\documentclass{article}
\RequirePackage[a4paper]{geometry}
\usepackage{mathptmx,graphicx}
\usepackage[dvipsnames]{xcolor}
\usepackage{abstract} 
\usepackage[numbers,sort&compress]{natbib}
\usepackage{hyperref}
\usepackage[labelfont=bf]{caption}
\usepackage{indentfirst}
\usepackage{amssymb}
\usepackage{amsmath}
\usepackage{diagbox}
\usepackage{svg}
\begin{document}
\begin{center}
{\Large\bfseries
Toward Micro-Endoscopy: Distal-Free, Configuration-Agnostic Focusing Through Multimode Fiber \par}
\vspace{3ex}
{\bfseries
Dvir Marsh$^{1}$, Lior Fridman$^{1}$, Stav Lotan$^{1}$, Amit Kam$^{2}$, Shie Mannor$^{1}$, Guy Bartal$^{1}$\par}
{\footnotesize\itshape
1. Andrew and Erna Viterbi Department of Electrical \& Computer Engineering, Technion, Haifa 320003, Israel\\
}
{\footnotesize\itshape
2. Department of Physics, Technion - Israel Institute of Technology, Haifa 32000, Israel\\
}
\vspace{3ex}
\end{center}

\begin{abstract}
\textbf{
Multimode fibers (MMFs) can transmit multiple guided modes simultaneously, making them a promising platform for high-resolution biomedical imaging, endoscopy and high-bandwidth optical communication. However, their complex modal behavior, influenced by environmental perturbations and mode coupling, presents a major challenge for accurate wavefront control. Conventional approaches for shaping the light at their output typically rely on the transmitted field as a source for iterative feedback, making it impractical for in-situ applications where direct access to the transmission is impossible. Here, we introduce a deep learning-based framework for predicting transmission through MMF by observing only the reflected signal. Harnessing the reflected signals that encode the fiber's internal configuration, our approach not only generalizes across varying fiber conditions but also enables focusing through the fiber without requiring transmission feedback. By training the system experimentally using a dataset of 4 million images across 1200 distinct fiber configurations, we demonstrate robust and precise wavefront reconstruction even under significant perturbations. Our results underscore the potential of learning-based techniques for real-time MMF-based imaging and optical communications, paving the way for efficient non-invasive focusing in practical applications.
}
\end{abstract}

\section{Introduction}
Multimode fibers (MMFs) have gained significant attention due to their potential to revolutionize various optical applications ~\cite{applications_review, single_multi}. Unlike single-mode fibers, which guide light through a single propagation mode, MMFs support the transmission of multiple spatial modes simultaneously ~\cite{Saleh}, with each mode effectively functioning as an independent data point or communication channel ~\cite{communication_channels}. This property makes MMFs highly attractive for minimally invasive deep-tissue endoscopy ~\cite{endoscopy_example, scanner_free, STABLE}. The transition from multi-core to multimode endoscopes enables significantly  smaller, more flexible endoscopes, reducing the probe diameter by 10x and more while maintaining imaging capacity ~\cite{resolution_comparison}. This miniaturization minimizes tissue damage and unlocks optical access to confined biological regions that remain completely unreachable by traditional probes ~\cite{imaging_example}. Furthermore, MMFs could also be attractive for high-capacity optical communication ~\cite{multiplexing_example, high_speed_transmission}, where mode-division multiplexing can enhance bandwidth for next-generation telecommunication systems.
Despite their advantages, MMFs present significant challenges due to the complex nature of modal propagation. In an idealized scenario, the eigenmodes of a perfectly uniform fiber are well defined, allowing for analytical wave propagation calculations ~\cite{fiber_book}. However, in real-world conditions, variations in the fiber fabrication process, external perturbations such as temperature fluctuations, fiber bending, and mechanical stress induce mode coupling ~\cite{mode_coupling, optical_power_flow}. This coupling forces energy to exchange unpredictably between these nominally independent eigenmodes, leading to severe distortions of the transmitted field ~\cite{seeing_through_chaos, learning_and_avoiding}. Consequently, the information carried by the individual modes is effectively scrambled, degrading the image quality. 

This poses a major obstacle for precise wavefront prediction, necessitating accurate modeling of the MMF transmission to account for the coupled eigenmodes. Conventional analytical approaches often fail due to the inherent complexity of the mode mixing. To illustrate this complexity, a $50 [\mu m]$ fiber supports $\sim 1500$ modes, consequently, characterizing the fiber's configuration requires mapping a complex transmission matrix with over two million degrees of freedom. Even microscopic physical perturbations result in unpredictable distortions, completely scrambling these phase relations into a new, unknown configuration within this massive space.

Wavefront shaping is a powerful approach to address the challenges of complex mode propagation in MMFs, offering strategies to control and optimize the transmitted field despite the modal scrambling. Two main strategies have emerged within this framework, both relying on spatial light modulators (SLMs) to shape the input wavefront. The first strategy employs a closed-loop feedback mechanism, where the input is iteratively adjusted to achieve a desired output using feedback from the transmitted field
~\cite{genetic_algorithm, endoscopy_review,phase_control}. The second is the transmission matrix (TM) approach, which characterizes the full input-output relationship of the fiber by measuring its complex-valued TM~\cite{TM_Gigan}. Once obtained, the TM enables deterministic control over the output field, allowing for precise focusing and imaging ~\cite{TM_applications, mode_control}. While both strategies have shown strong performance in controlled settings, they both typically require access to the transmitted field and are highly sensitive to environmental perturbations like fiber bending and thermal fluctuations. This makes them less practical for in-situ applications such as endoscopy. Although recent advancements have achieved distal-free focusing in biological tissues by exploiting the tilt-shift memory effect ~\cite{rapid_wavefront}, these techniques are fundamentally inapplicable to MMFs, which lack this specific memory effect and severely scramble polarization states, thereby violating the necessary scalar field approximations.

A promising alternative to conventional wavefront shaping methods is to harness deep learning (DL), wherein Deep Neural Networks (DNNs) learn the complex mapping between input and output fields, enabling  the prediction of light propagation through MMFs without relying on direct output measurements~\cite{light_propagation_prediction, imaging_through_multimode_fibers, transmission_of_natural_scene_images, learning_to_see, image_reconstruction}. Once trained, DNNs perform inference orders of magnitude faster than iterative TM-reconstruction or closed-loop optimization, enabling real-time or high-frame-rate imaging and focusing through MMFs for live endoscopy and sensing  ~\cite{all_fiber_high_speed}. Conventional DNNs rely solely on the input wavefront, lacking any conditional information regarding the instantaneous fiber state. Consequently, a network trained on a static configuration will predictably fail to generalize when the fiber is subjected to mechanical perturbations. This severely limits the robustness and practical applicability of standard DNNs for predicting MMF transmission in dynamic, uncontrolled environments.

Here, we present a robust, high-contrast focusing through an MMF in arbitrary and unknown configurations, overcoming the need for distal or fluorescent feedback ~\cite{Anats_work}. We train a DNN on paired transmitted and reflected measurements collected across numerous input patterns and fiber configurations, enabling accurate prediction of the transmission using the input and reflected signals. This approach paves the way to an adaptive solution for various MMF-based applications in photonics and biomedical imaging, where direct access to the distal facet of the fiber is unavailable.

\section{Results}
\paragraph*{The principle} The behavior of light propagating through an MMF can be comprehensively described by its complex-valued TM, denoted as $T\in \mathbb{C}^{N\times N}$, where N represents the number of guided modes supported by the fiber. This matrix linearly maps the complex mode distribution $u$ of the input field in the fiber's input plane to the complex mode distribution $v$ in the output plane, i.e. $v=T\cdot u$; correspondingly, the reflection matrix (RM), $R$, maps $u$ to the reflected wave $w=R\cdot u$.

Ideally, a perfectly uniform fiber supports eigenmodes that travel independently, yielding a purely diagonal TM ~\cite{seeing_through_chaos}. In practice, however, mechanical perturbations inevitably induce coupling between these modes ~\cite{MMF_memory_effect}, destroying the matrix's diagonality thus making the TM  dependent on the fiber configuration.

Since RM, $R$, and TM, $T$, are related to the same underlying fiber configuration, the two are linked by a complex deterministic mapping ~\cite{Light_fields_in_complex_media}. Namely, the information about the fiber configuration, encoded in the reflected signal, can, in principle, be utilized to predict the transmission. 

\paragraph*{Reflection-Informed Deep Learning} To exploit this physical coupling, we designed a reflection-informed DNN based on a U-Net architecture ~\cite{U_Net}. The network is trained to map a set of measurements - specifically, the phase pattern applied by a Spatial Light Modulator (SLM) and the resulting reflected intensities in both the image and Fourier planes - to the transmitted speckle pattern at the distal end (Fig. \ref{fig: model}). By training this model across thousands of randomly induced fiber configurations, the network learns to implicitly decode the fiber's instantaneous state from the reflection, estimating the transmission without direct output measurements. Full architectural and training details are provided in the Methods section.

\begin{figure}[h!]
    \centering
    \includegraphics[width=1.0\linewidth]{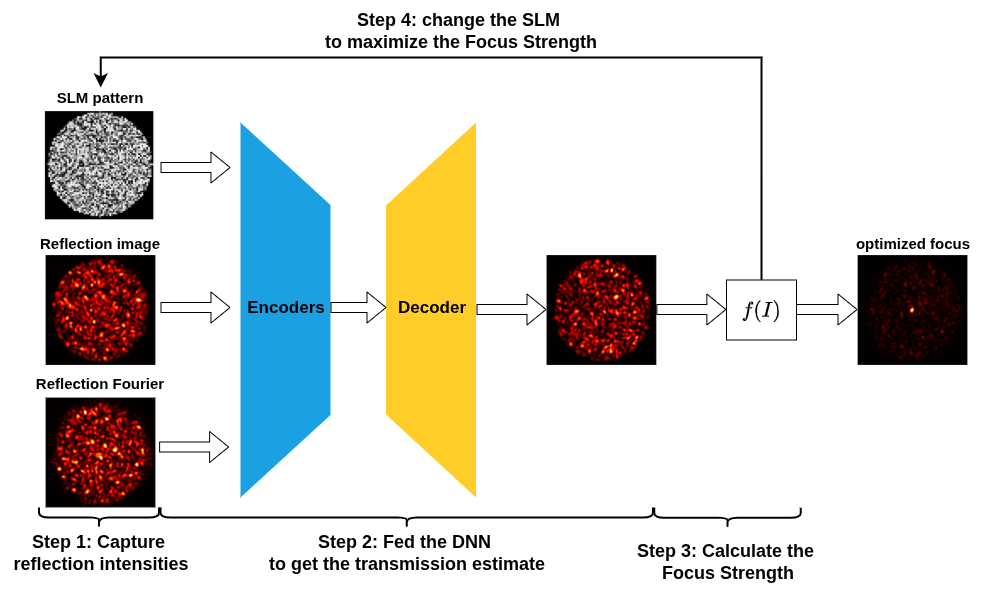}
    \caption{\textbf{Schematic of the indirect feedback focusing process. } The procedure is as follows: \textbf{1.} The reflected intensities are measured. \textbf{2.} These measurements, along with the current SLM pattern, are fed into the DNN to obtain a transmission estimate. \textbf{3.} The FS metric (Eq. ~\ref{focus_strength}) is calculated from the estimated transmission. \textbf{4.} A genetic algorithm optimizes the SLM phase pattern based on this metric, iterating without any direct feedback. }
    \label{fig:focus_flow}
\end{figure}

\paragraph*{Distal-Free Focusing} We validate our model's practical utility by performing indirect-feedback focusing, guiding an input wavefront optimization process solely based on the model's transmission predictions. This process, illustrated in Fig. \ref{fig:focus_flow}, involves measuring the reflected signals, feeding them into the trained model to estimate the transmitted field, and calculating the Focus Strength (FS) from this prediction. The FS (Eq.~\ref{focus_strength}) - defined as the ratio of intensity within a target focal region to the average background intensity - serves as the objective metric to guide the optimization algorithm for the next SLM phase update.

This experiment was repeated for 100 independent focusing processes where, for each process, the fiber was placed in a random, unknown configuration unseen during the network's training. For each optimization step, we simultaneously captured the ground-truth (GT) transmitted field using the transmission camera to calculate the GT FS, allowing us to evaluate the model's accuracy. An example of this iterative process is shown in Supplementary Video 1.

Fig. \ref{fig:focus_process} presents the average FS and its standard deviation over the 100 processes as a function of the optimization step. As shown by the blue curve, the ground-truth FS measured by the camera increases monotonically and converges to a high-intensity focus. This demonstrates that optimizing the SLM pattern based on the model's predictions effectively forces the real, physical transmitted field to focus. This confirms that our framework successfully enables robust, indirect feedback focusing through MMFs across unknown fiber configurations.

\begin{figure}[h!]
    \centering
    \includegraphics[width=1.0\linewidth]{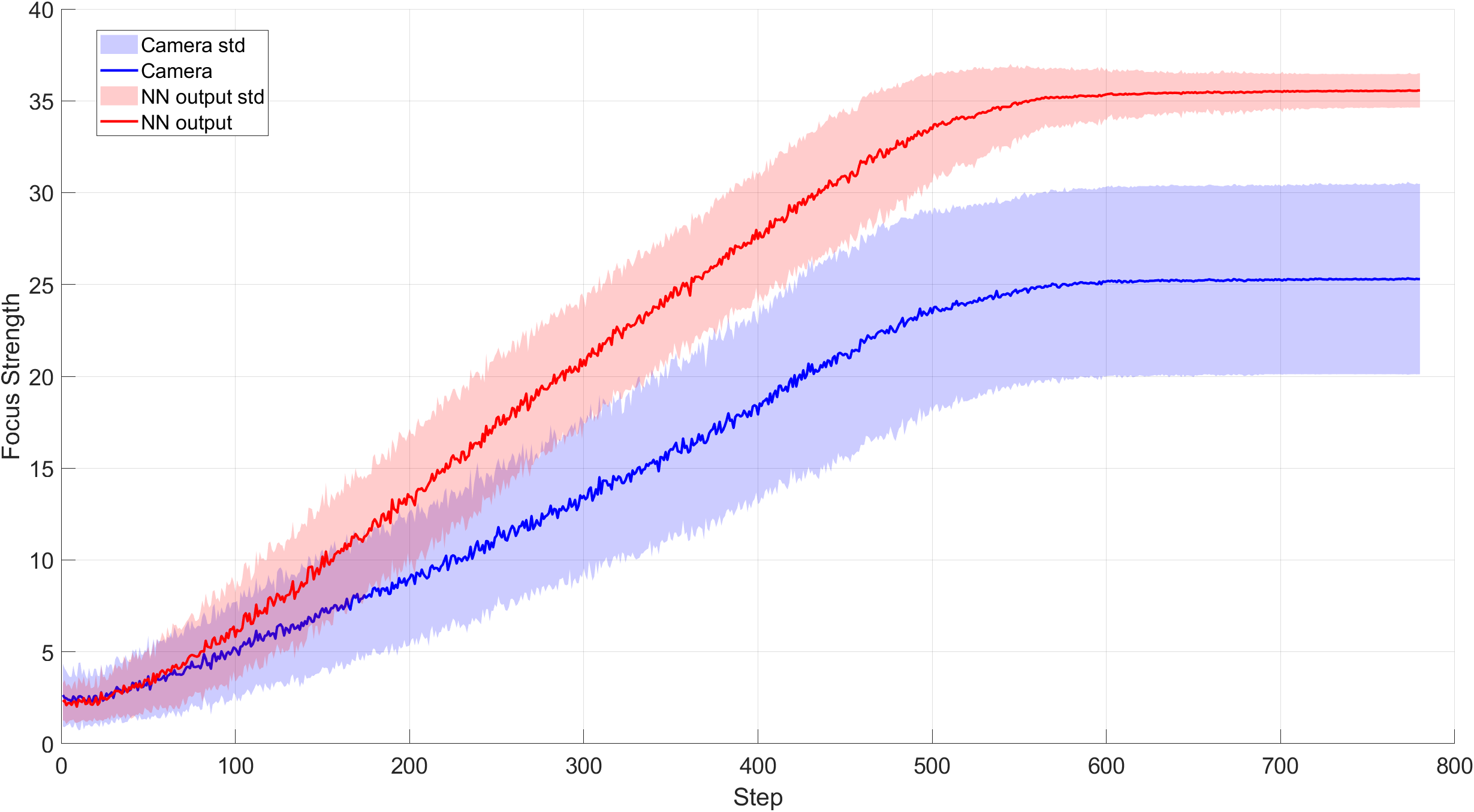}
    \caption{\textbf{Average FS during indirect feedback optimization. } The plot shows the average FS (solid lines) and standard deviation (shaded areas) over 100 independent focusing processes. The FS calculated from the model's prediction, which served as the feedback signal for the optimization algorithm, is shown in red. The corresponding ground-truth FS, calculated from the camera on the transmission side, is shown in blue. The monotonic increase in the GT (blue) curve demonstrates that optimizing based on the model's predictions effectively leads to a strong focus in the real transmitted field, validating the indirect feedback approach.}
    \label{fig:focus_process}
\end{figure}

\paragraph*{Model Performance} We further evaluate our reflection-informed DNN model performance based on its ability to accurately predict the transmitted field through the MMF across a range of unseen fiber configurations. The dataset was split such that the model was trained on 960 fiber configurations, while the remaining 240 configurations were reserved for validation and the test set. The evaluation focused on both the qualitative fidelity of the predicted speckle patterns and the quantitative accuracy over various metrics.

\begin{figure}[h!]
\centerline{\includegraphics[width=1.0\columnwidth]{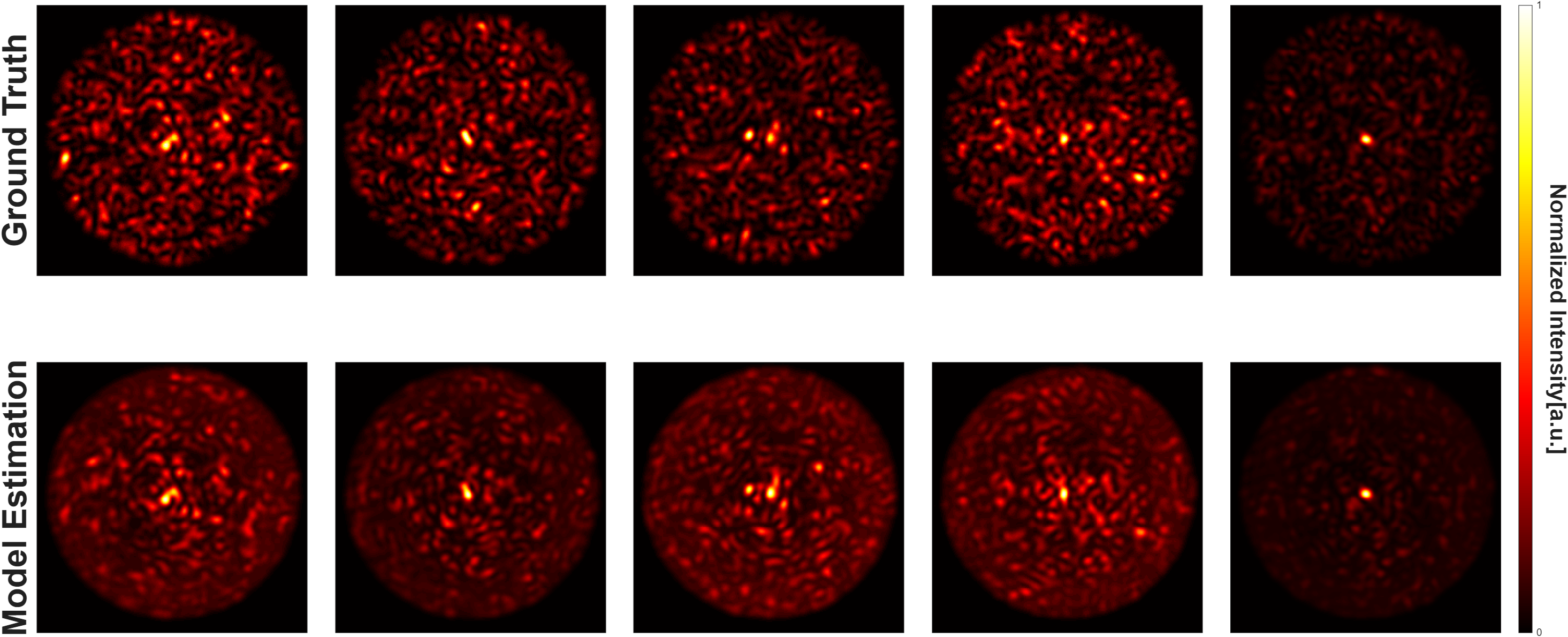}}
\caption{\textbf{Qualitative comparison of predicted and measured fields.} A side-by-side comparison of fiber configurations that are not included in the training dataset. The \textbf{Top row} presents the GT transmitted speckle patterns measured by the camera. The \textbf{Bottom row} shows the corresponding speckle patterns predicted by our DNN model without any transmission measurement. The corresponding FS is presented above each image.}
\label{fig: example}
\end{figure}

A qualitative comparison between the experimentally GT transmitted fields and the corresponding fields predicted by our model (for a set of fiber configurations outside the training dataset) is shown in Fig.~\ref{fig: example}. The model successfully captures the salient features and overall structure of the complex speckle patterns generated at the MMF output, even under significant mode coupling induced by the mechanical perturbations. Although minor discrepancies exist, the visual similarity between the ground truth and predicted patterns indicates that the network has learned a robust mapping from the input wavefront and reflected signals to the transmitted field despite the inherent complexity and variability of MMF propagation.

To provide a quantitative measure of the model's predictive capability, particularly in the context of focusing applications, we utilize a FS metric, which is a variation of the enhancement factor ~\cite{focusing_coherent}. We define the FS, $f(I)$, for a given intensity distribution $I(x,y)$ as the ratio of the average intensity within a target focal region (a small, diffraction-limited area centered on the desired focal point) to the average intensity over the entire image area:
\begin{equation}
    f(I)=\frac{\frac{1}{M}\sum\limits_{(x,y)\in S}\left[I(x,y)\right]}{\frac{1}{N}\sum\limits_{x,y}\left[I(x,y)\right]}
    \label{focus_strength}
\end{equation}
Here, $S=(x_0-w,x_0+w)\times (y_0-w,y_0+w)$ is the focal region, $(x_0,y_0)$ denotes the coordinates of the target focal point, $w$ is half the diffraction limit, and M and N are the number of pixels in the focal region and the entire image, respectively. A higher FS indicates a greater concentration of light energy in the desired spot.

We calculated the FS for both the experimentally measured transmitted images ($I$) and the images predicted by the model ($\hat{I}$) for each test case. To quantify the prediction error, we defined the focus deviation (FD) as the average relative difference between the FS calculated from the ground truth and the model's prediction:
\begin{equation}
    FD\left(\mathcal{D}\right)\overset{\Delta}{=}\mathbb{E}_{\mathcal{D}}\left[\frac{\left|f(I)-f(\hat{I})\right|}{f(I)}\right]
\end{equation}

The expectation value $\mathbb{E}_\mathcal{D}[\cdot]$ is taken over a specific dataset $\mathcal{D}$, e.g., training or test dataset. Our evaluation revealed an average focus deviation of only 14\% across the 240 test fiber configurations (test and validation). This small discrepancy is a strong indicator of the model's robust generalization, confirming that it can reliably predict transmission - and thus enable high contrast focusing - even when the physical configuration of the MMF is unknown.

To further quantify the model's advantage, we compare its performance against a baseline UNet model that was trained using only the SLM patterns as input, without the reflection information (for fairness, both models were trained with the same number of parameters). To emphasize the superiority of our reflection-informed approach under more significant perturbations, the models compared here were trained and tested on a dataset where the fiber was subjected to a wider range of mechanical bending via the motorized stage. Table \ref{tab:placeholder_label} summarizes the performance of both models on the unseen test set across several standard metrics. Our reflection-informed model demonstrates superior performance on the test set across all fronts: it achieves a higher Structural Similarity Index Measure (SSIM) and Pearson correlation coefficient, indicating greater fidelity in reconstructing the overall structure and pixel-wise values of the speckle patterns. Crucially, it results in a significantly lower Focus Deviation and Mean Squared Error (MSE), confirming that incorporating the reflected signal provides critical information about the fiber's configuration, leading to more accurate and reliable transmission predictions.

\begin{table}
    \centering
    \caption{\textbf{Quantitative performance comparison.} The table contrasts our reflection-informed model with a baseline using only the SLM input. Our model shows superior performance across all metrics on the test set: Structural Similarity Index (SSIM), Pearson correlation, Focus Deviation, and Mean Squared Error (MSE).}
    \label{tab:placeholder_label}
    \begin{tabular}{|c|c|c|}\hline
         \backslashbox{Metric}{Model}&  Reflection informed UNet& SLM only UNet\\\hline
         SSIM&  $0.974\pm 0.005$& $0.958\pm 0.006$\\\hline
         Pearson correlation&  $0.82\pm 0.026$& $0.78\pm 0.05$\\\hline
         Focus deviation&  $17.07\%\pm 7.25\%$& $22.8\%\pm 12.9\%$\\\hline
         MSE&  $12.27\pm 0.79$& $14.83\pm 1.07$\\ \hline
    \end{tabular} 
\end{table}

\begin{figure}[h!]
\centerline{\includegraphics[width=1.0\columnwidth]{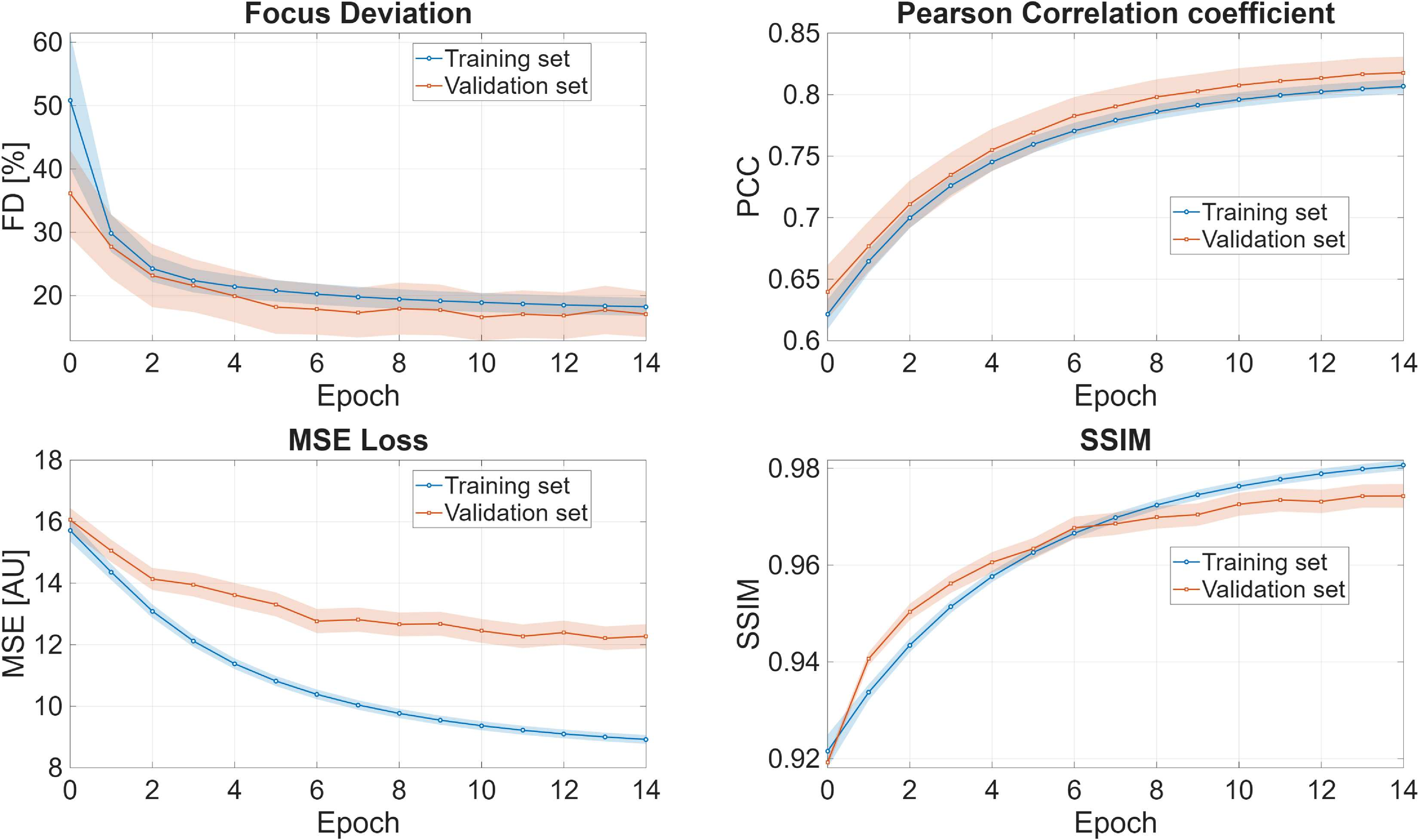}}
\caption{\textbf{Convergence diagnostics over training.} Each panel compares training (blue) and validation (orange) traces for (top left) Loss, (top right) SSIM, (bottom left) Focus Deviation, and (bottom right) Pearson correlation coefficient. The plot shows the average metrics (solid lines) and their standard deviation (shaded areas).}
\label{fig:convergence_plot}
\end{figure}

Those networks were trained for 15 epochs using the Adam optimizer, with early stopping triggered after 3 epochs of no improvement. Figure \ref{fig:convergence_plot} shows the training and validation traces for four metrics of the reflection-informed model: Loss, SSIM, Focus Deviation, and Pearson correlation. Note that this plot is taken from a training run on a dataset with a wider motor range. Over the run, the training loss falls continuously; the validation loss also decreases but at a slower rate and remains consistently higher than the training loss, indicating a persistent generalization gap despite overall improvement. Concurrently, SSIM and the Pearson correlation steadily increase and begin to saturate, indicating improved reconstruction fidelity and stronger correspondence to the ground truth. Validation Focus Deviation decreases (improves) across epochs. Overall, the combined qualitative and quantitative metrics demonstrate reliable convergence, although the remaining validation–training gap suggests there may still be benefit from further regularization, data augmentation, or additional training data.

\section{Methods}
\begin{figure}[h]
\centerline{\includegraphics[width=1.0\columnwidth]{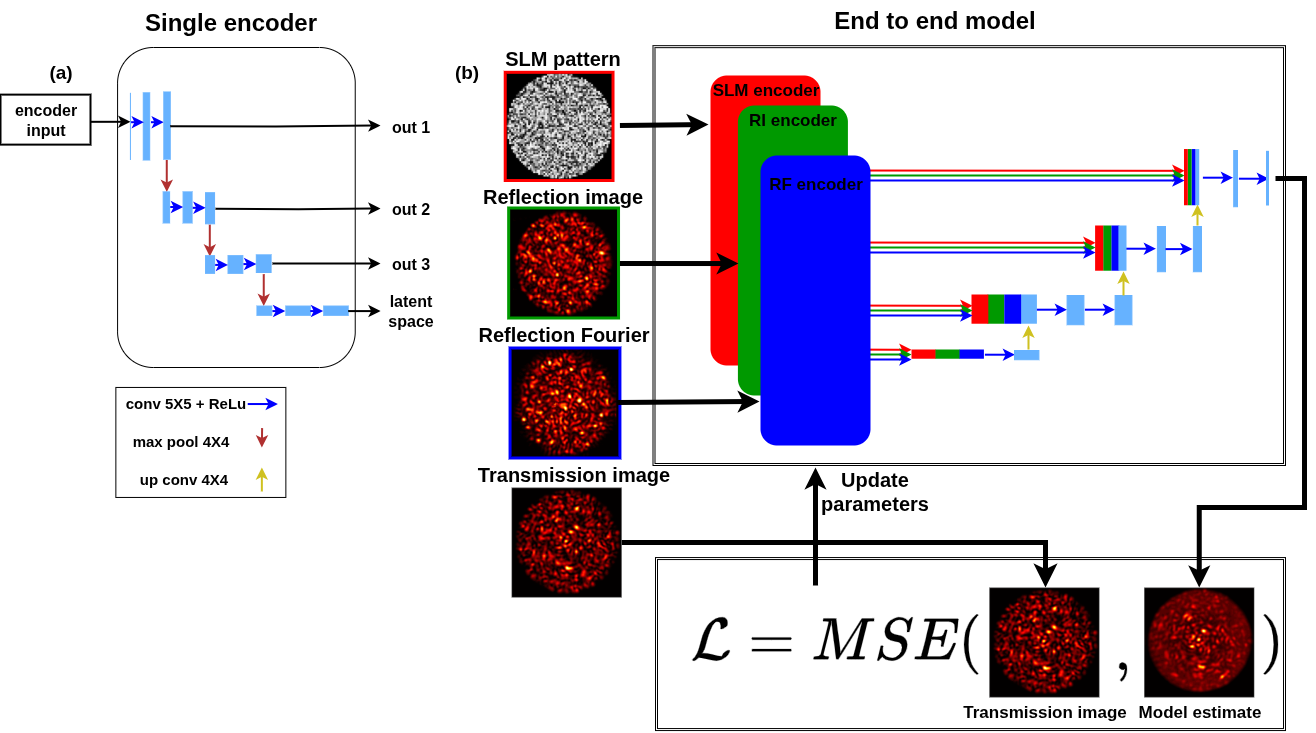}}
\caption{\textbf{Reflection-informed DNN architecture} The network estimates the transmitted intensity in the image plane based on three input modalities: (1) the SLM pattern, (2) the reflected intensity measured in the image plane (RI), and (3) the reflected intensity measured in the Fourier plane (RF). Each input modality is fed into its own encoder (SLM, RI and RF encoders); the three encoders share the same structure, (a) illustration of the encoder architecture. (b)  the full end-to-end model: encoder outputs at each resolution level are forwarded to the decoder via U-Net style skip connections. Unlike a single-encoder U-Net, the feature maps produced by all three encoders are concatenated together at each decoder level before the up-convolutions. The network produces an estimate of the transmission intensity in the image plane.
}
\label{fig: model}
\end{figure}

\paragraph*{The DNN Model} Our DNN, presented in Fig.~\ref{fig: model}, features three separate encoder branches to process the three input modalities: the pattern launched to the fiber, generated by SLM - a device capable of shaping the spatial phase of an incident wave;  the intensity of the electromagnetic field launched back from the fiber input after propagating back and forth in the fiber ('reflected signal'); and the intensity in the Fourier plane of this reflected signal. The output feature vectors from these encoders are concatenated and fed into a single decoder, which uses a series of up-convolutional layers to reconstruct the estimated intensity distribution of the transmitted field. We create a dataset that is randomly split such that the model is trained on data from specific fiber configurations and validated on data from unseen fiber configurations. The model is trained by minimizing the mean squared error (MSE) loss:
\begin{equation}
\mathcal{L}=\frac{1}{M}\sum\limits_{i=1}^{M}(I_i-\hat{I}_i)^2
\label{MSE_loss}
\end{equation}
Here, $M$ is the total number of pixels in the image, while $I_i$ and $\hat{I}_i$ are the intensity values at the i-th pixel in the ground-truth and predicted images, respectively. The loss value, $\mathcal{L}$, quantifies the discrepancy between the model's prediction and the actual measurement. During training, the network's parameters are iteratively updated to minimize this loss, thereby learning the complex mapping from the input wavefront and reflected fields to the transmitted speckle pattern.

\paragraph*{The Dataset} This DNN is trained by a comprehensive dataset that consists of 4.5 million sets of images collected from 1200 distinct fiber configurations. Each configuration contains 10 independent focusing processes using a genetic algorithm ~\cite{genetic_algorithm}, which iteratively adjusts the SLM phase mask to maximize the focus strength (FS) metric (Eq.~\ref{focus_strength}). During each step of this optimization, we captured a complete set of images from all four cameras (transmitted image/Fourier plane, reflected image/Fourier plane) and recorded the corresponding SLM phase mask. Each image was preprocessed by cropping the relevant speckle pattern and resizing it to $256\times256$ pixels, yielding a rich collection of input-output pairs. 

\begin{figure}[h!]
\centerline{\includegraphics[width=1.0\columnwidth]{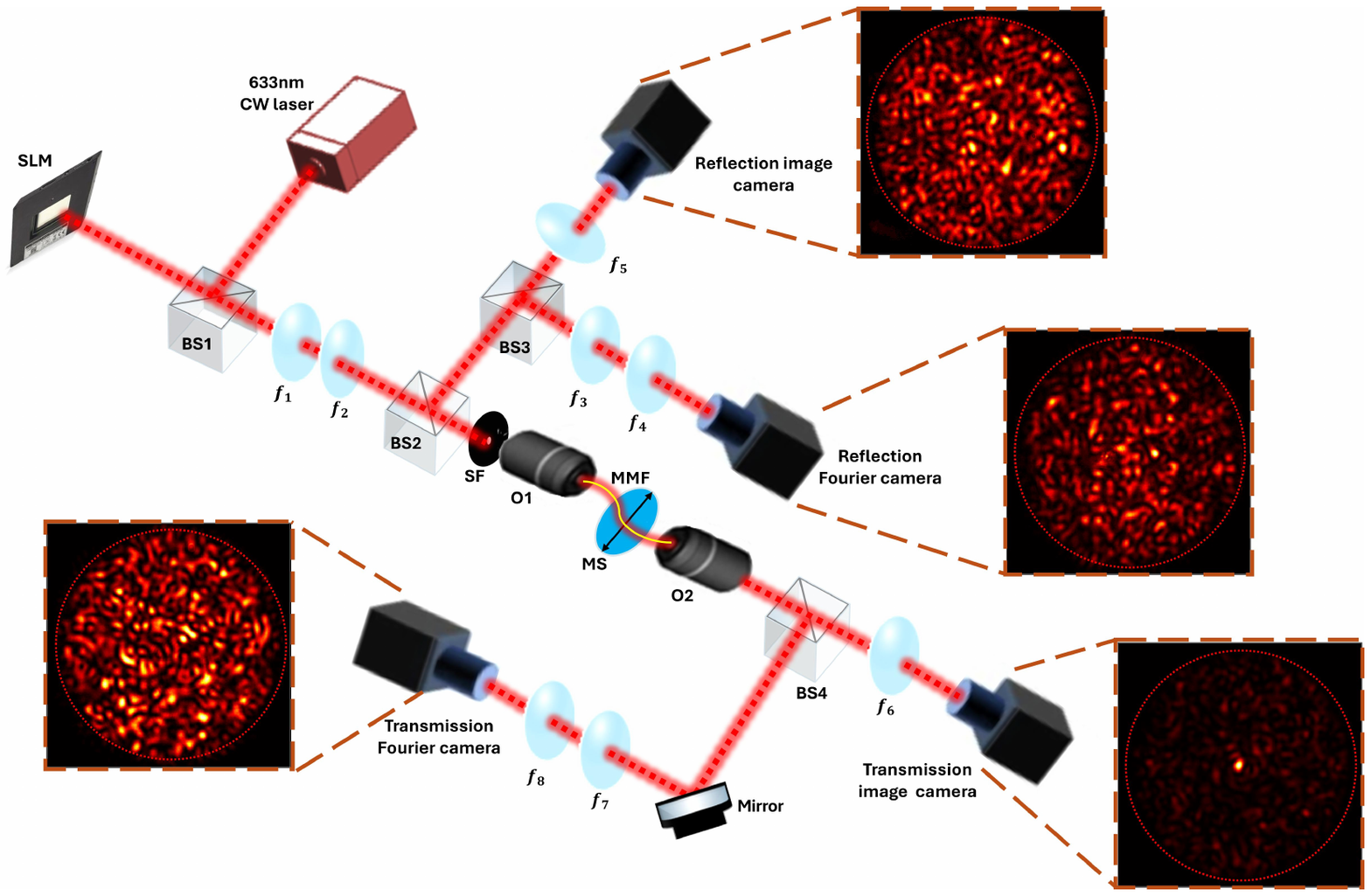}}
\caption{\textbf{Experimental setup for data acquisition.}  A collimated HeNe laser beam ($\lambda=633nm$) is modulated by an SLM, whose facet is imaged with a 4F system (with $f_1, f_2$ lenses), and coupled into the MMF by an objective lens (O1). The transmitted and reflected fields are simultaneously captured by separate cameras in both the image and Fourier planes; an example for each is displayed in the inset. An angled fiber input facet and a spatial filter (SF) are used to isolate  the useful reflection signal from the unwanted reflections.  A motorized stage (MS) with a range of motion of 12$mm$ was introduced for controlled variations in fiber configurations.}
\label{fig: Exp}
\end{figure}

\paragraph*{The Experimental Setup} This dataset was acquired using an experimental setup designed to simultaneously control the input wavefront and capture both transmitted and reflected fields under various controlled fiber perturbations. The setup, illustrated in Fig.~\ref{fig: Exp}, used a collimated 633-nm Helium-Neon (HeNe) laser beam as the light source. The SLM was used to precisely control the input wavefront, $u$, by imposing a spatially varying phase pattern onto the beam. The modulated light was then relayed through a 4F imaging system and coupled into the MMF (50 $\mu m$ core diameter, supporting $\sim1500$ modes) using a microscope objective lens (50x, NA 0.42). For both the transmitted and reflected paths, we recorded intensity distributions in both the image plane and the Fourier plane using separate cameras, as the combination of these provides a more complete representation of the complex optical field. An angled fiber-input facet in combination with a spatial filter was employed to isolate two distinct reflections: (i) the direct angled reflection from the proximal facet, which contains no information about the MMF as propagation has not yet occurred, and (ii) the weaker reflection from the distal facet, which propagates back through the proximal facet without an angle and conveys information about propagation and internal reflections within the MMF. To induce different environmental conditions, the fiber was mounted on a motorized stage with a 12 mm range of motion, allowing us to apply controlled mechanical bending and generate distinct fiber configurations.

\section{Discussion}
In this work, we demonstrated a novel framework for indirect focusing through multimode fibers, successfully decoupling the wavefront optimization process from distal facet feedback. By leveraging a reflection-informed DNN, we showed that the back-reflected signal encodes sufficient information to accurately predict the transmitted field across strictly unseen fiber states. The model achieved an average focus deviation of only 14\% across 240 test configurations, confirming its ability to reliably interpolate within the high-dimensional perturbation space sampled during training. By mitigating the traditional reliance on transmission-side access, this approach represents a major step toward deploying MMFs in dynamic, real-world environments.

As evidenced by our quantitative comparison, the reflection-informed DNN consistently outperformed the baseline SLM-only network across all metrics. Since the baseline model is fundamentally blind to the fiber's instantaneous state, we anticipate that this performance gap will widen significantly under greater mechanical perturbations. As the physical deformation increases, the mapping from the proximal input to the distal output becomes increasingly unpredictable without the conditional guidance provided by the back-reflected signal.

Translating this framework to clinical settings presents two clear scaling trajectories. First, utilizing larger core fibers will necessitate higher-capacity network architectures to handle the quadratically expanding transmission matrices. Second, deploying this technology in highly torturous clinical pathways requires extending the model's extrapolative robustness. As standard DNNs inherently struggle to extrapolate beyond their training distributions ~\cite{extrapolation_limitation, extrapolation_and_interpolation}, future architectures could integrate Physics-Informed Neural Networks (PINNs) ~\cite{PINN, PINN_extrapolation}. Embedding wave-propagation equations directly into the loss function would constrain the model to learn the underlying physics, potentially facilitating reliable transmission estimates under extreme, previously unseen clinical bending.

Finally, moving beyond iterative algorithms represents the next major milestone for practical deployment. Recent breakthroughs have achieved video-rate imaging through dynamically bent multi-core fibers by adding partially reflecting mirrors to the distal tip ~\cite{Badt_Katz_2022}. However, realizing single-shot control in multimode fibers without modifying the distal facet remains a significant challenge. While our current genetic algorithm operates entirely without distal feedback, the iterative nature of the optimization inherently limits the focusing speed. Future work could bypass this bottleneck entirely by training an inverse-design network to directly map the measured reflection to the required proximal SLM phase mask, paving the way for true single-shot, real-time focusing through dynamic optical fibers.

\bibliography{references}

\begin{thebibliography}{10}

\bibitem{applications_review}
Hui Cao, Tomáš Čižmár, Sergey Turtaev, Tomáš Tyc, and Stefan Rotter.
\newblock Controlling light propagation in multimode fibers for imaging, spectroscopy, and beyond.
\newblock {\em Advances in Optics and Photonics}, 15(2):524--612, June 2023.

\bibitem{single_multi}
Qiang Wu, Yuwei Qu, Juan Liu, Jinhui Yuan, Sheng-Peng Wan, Tao Wu, Xing-Dao He, Bin Liu, Dejun Liu, Youqiao Ma, Yuliya Semenova, Pengfei Wang, Xiangjun Xin, and Gerald Farrell.
\newblock Singlemode-{Multimode}-{Singlemode} {Fiber} {Structures} for {Sensing} {Applications}—{A} {Review}.
\newblock {\em IEEE Sensors Journal}, 21(11):12734--12751, June 2021.

\bibitem{Saleh}
Fiber {Optics}.
\newblock In {\em Fundamentals of {Photonics}}, pages 272--309. John Wiley \& Sons, Ltd, 1991.
\newblock Section: 8 \_eprint: https://onlinelibrary.wiley.com/doi/pdf/10.1002/0471213748.ch8.

\bibitem{communication_channels}
Alireza Tarighat, Rick~C.J. Hsu, Akhil Shah, Ali~H. Sayed, and Bahram Jalali.
\newblock Fundamentals and challenges of optical multiple-input multiple-output multimode fiber links [{Topics} in {Optical} {Communications}].
\newblock {\em IEEE Communications Magazine}, 45(5):57--63, May 2007.

\bibitem{endoscopy_example}
Elizabeth Abraham, Junxiao Zhou, and Zhaowei Liu.
\newblock Speckle structured illumination endoscopy with enhanced resolution at wide field of view and depth of field.
\newblock {\em Opto-Electronic Advances}, 6(7):220163--8, July 2023.

\bibitem{scanner_free}
Youngwoon Choi, Changhyeong Yoon, Moonseok Kim, Taeseok~Daniel Yang, Christopher Fang-Yen, Ramachandra~R. Dasari, Kyoung~Jin Lee, and Wonshik Choi.
\newblock Scanner-{Free} and {Wide}-{Field} {Endoscopic} {Imaging} by {Using} a {Single} {Multimode} {Optical} {Fiber}.
\newblock {\em Physical Review Letters}, 109(20):203901, November 2012.

\bibitem{STABLE}
Zhong Wen, Zhenyu Dong, Qilin Deng, Chenlei Pang, Clemens~F. Kaminski, Xiaorong Xu, Huihui Yan, Liqiang Wang, Songguo Liu, Jianbin Tang, Wei Chen, Xu~Liu, and Qing Yang.
\newblock Single multimode fibre for in vivo light-field-encoded endoscopic imaging.
\newblock {\em Nature Photonics}, 17(8):679--687, August 2023.

\bibitem{resolution_comparison}
Demetri Psaltis and Christophe Moser.
\newblock Imaging with multimode fibers.
\newblock {\em Optics and Photonics News}, 27.1, 2016.

\bibitem{imaging_example}
Tomáš Čižmár and Kishan Dholakia.
\newblock Exploiting multimode waveguides for pure fibre-based imaging.
\newblock {\em Nature Communications}, 3(1):1027, August 2012.

\bibitem{multiplexing_example}
D.~J. Richardson, J.~M. Fini, and L.~E. Nelson.
\newblock Space-division multiplexing in optical fibres.
\newblock {\em Nature Photonics}, 7(5):354--362, May 2013.

\bibitem{high_speed_transmission}
Ronald~E. Freund, Christian-A. Bunge, Nikolay~N. Ledentsov, D.~Molin, and Ch. Caspar.
\newblock High-{Speed} {Transmission} in {Multimode} {Fibers}.
\newblock {\em Journal of Lightwave Technology}, 28(4):569--586, February 2010.

\bibitem{fiber_book}
Katsunari Okamoto.
\newblock Chapter 3 - {Optical} fibers.
\newblock In Katsunari Okamoto, editor, {\em Fundamentals of {Optical} {Waveguides} ({Second} {Edition})}, pages 57--158. Academic Press, Burlington, January 2006.

\bibitem{mode_coupling}
Luca Palmieri.
\newblock Coupling mechanism in multimode fibers.
\newblock In {\em Next-{Generation} {Optical} {Communication}: {Components}, {Sub}-{Systems}, and {Systems} {III}}, volume 9009, pages 87--95. SPIE, February 2014.

\bibitem{optical_power_flow}
D.~Gloge.
\newblock Optical {Power} {Flow} in {Multimode} {Fibers}.
\newblock {\em Bell System Technical Journal}, 51(8):1767--1783, 1972.
\newblock \_eprint: https://onlinelibrary.wiley.com/doi/pdf/10.1002/j.1538-7305.1972.tb02682.x.

\bibitem{seeing_through_chaos}
Martin Plöschner, Tomáš Tyc, and Tomáš Čižmár.
\newblock Seeing through chaos in multimode fibres.
\newblock {\em Nature Photonics}, 9(8):529--535, August 2015.

\bibitem{learning_and_avoiding}
Maxime~W. Matthès, Yaron Bromberg, Julien De~Rosny, and Sébastien~M. Popoff.
\newblock Learning and {Avoiding} {Disorder} in {Multimode} {Fibers}.
\newblock {\em Physical Review X}, 11(2):021060, June 2021.

\bibitem{genetic_algorithm}
Donald~B. Conkey, Albert~N. Brown, Antonio~M. Caravaca-Aguirre, and Rafael Piestun.
\newblock Genetic algorithm optimization for focusing through turbid media in noisy environments.
\newblock {\em Optics Express}, 20(5):4840, February 2012.

\bibitem{endoscopy_review}
Guangxing Wu, Runze Zhu, Yanqing Lu, Minghui Hong, and Fei Xu.
\newblock Optical scanning endoscope via a single multimode optical fiber.
\newblock {\em Opto-Electronic Science}, 3(3):230041--32, March 2024.

\bibitem{phase_control}
I.~M. Vellekoop and A.~P. Mosk.
\newblock Phase control algorithms for focusing light through turbid media.
\newblock {\em Optics Communications}, 281(11):3071--3080, June 2008.

\bibitem{TM_Gigan}
S.~M. Popoff, G.~Lerosey, R.~Carminati, M.~Fink, A.~C. Boccara, and S.~Gigan.
\newblock Measuring the {Transmission} {Matrix} in {Optics}: {An} {Approach} to the {Study} and {Control} of {Light} {Propagation} in {Disordered} {Media}.
\newblock {\em Physical Review Letters}, 104(10):100601, March 2010.

\bibitem{TM_applications}
Moonseok Kim, Wonjun Choi, Youngwoon Choi, Changhyeong Yoon, and Wonshik Choi.
\newblock Transmission matrix of a scattering medium and its applications in biophotonics.
\newblock {\em Optics Express}, 23(10):12648--12668, May 2015.

\bibitem{mode_control}
Moussa N’Gom, Theodore~B. Norris, Eric Michielssen, and Raj~Rao Nadakuditi.
\newblock Mode control in a multimode fiber through acquiring its transmission matrix from a reference-less optical system.
\newblock {\em Optics Letters}, 43(3):419--422, February 2018.

\bibitem{rapid_wavefront}
Sagi Monin, Marina Alterman, and Anat Levin.
\newblock Rapid wavefront shaping using an optical gradient acquisition.
\newblock {\em Nature Communications}, 17(1):1537, January 2026.

\bibitem{light_propagation_prediction}
Pengfei Fan, Liang Deng, and Lei Su.
\newblock Light {Propagation} {Prediction} through {Multimode} {Optical} {Fibers} with a {Deep} {Neural} {Network}.
\newblock In {\em 2018 {IEEE} 3rd {Advanced} {Information} {Technology}, {Electronic} and {Automation} {Control} {Conference} ({IAEAC})}, pages 1080--1084, October 2018.

\bibitem{imaging_through_multimode_fibers}
Eirini Kakkava, Babak Rahmani, Navid Borhani, Uğur Teğin, Damien Loterie, Georgia Konstantinou, Christophe Moser, and Demetri Psaltis.
\newblock Imaging through multimode fibers using deep learning: {The} effects of intensity versus holographic recording of the speckle pattern.
\newblock {\em Optical Fiber Technology}, 52:101985, November 2019.

\bibitem{transmission_of_natural_scene_images}
Piergiorgio Caramazza, Oisín Moran, Roderick Murray-Smith, and Daniele Faccio.
\newblock Transmission of natural scene images through a multimode fibre.
\newblock {\em Nature Communications}, 10(1):2029, May 2019.

\bibitem{learning_to_see}
Navid Borhani, Eirini Kakkava, Christophe Moser, and Demetri Psaltis.
\newblock Learning to see through multimode fibers.
\newblock {\em Optica}, 5(8):960--966, August 2018.

\bibitem{image_reconstruction}
Changyan Zhu, Eng~Aik Chan, You Wang, Weina Peng, Ruixiang Guo, Baile Zhang, Cesare Soci, and Yidong Chong.
\newblock Image reconstruction through a multimode fiber with a simple neural network architecture.
\newblock {\em Scientific Reports}, 11(1):896, January 2021.

\bibitem{all_fiber_high_speed}
Zhoutian Liu, Lele Wang, Yuan Meng, Tiantian He, Sifeng He, Yousi Yang, Liuyue Wang, Jiading Tian, Dan Li, Ping Yan, Mali Gong, Qiang Liu, and Qirong Xiao.
\newblock All-fiber high-speed image detection enabled by deep learning.
\newblock {\em Nature Communications}, 13(1):1433, March 2022.

\bibitem{Anats_work}
Dror Aizik, Ioannis Gkioulekas, and Anat Levin.
\newblock Fluorescent wavefront shaping using incoherent iterative phase conjugation.
\newblock {\em Optica}, 9(7):746--754, July 2022.

\bibitem{MMF_memory_effect}
Shuhui Li, Simon A.~R. Horsley, Tomáš Tyc, Tomáš Čižmár, and David~B. Phillips.
\newblock Memory effect assisted imaging through multimode optical fibres.
\newblock {\em Nature Communications}, 12(1):3751, June 2021.

\bibitem{Light_fields_in_complex_media}
Stefan Rotter and Sylvain Gigan.
\newblock Light fields in complex media: {Mesoscopic} scattering meets wave control.
\newblock {\em Reviews of Modern Physics}, 89(1):015005, March 2017.

\bibitem{U_Net}
Olaf Ronneberger, Philipp Fischer, and Thomas Brox.
\newblock U-{Net}: {Convolutional} {Networks} for {Biomedical} {Image} {Segmentation}, May 2015.
\newblock arXiv:1505.04597 [cs].

\bibitem{focusing_coherent}
I.~M. Vellekoop and A.~P. Mosk.
\newblock Focusing coherent light through opaque strongly scattering media.
\newblock {\em Optics Letters}, 32(16):2309--2311, August 2007.

\bibitem{extrapolation_limitation}
P.J. Haley and D.~Soloway.
\newblock Extrapolation limitations of multilayer feedforward neural networks.
\newblock In {\em [{Proceedings} 1992] {IJCNN} {International} {Joint} {Conference} on {Neural} {Networks}}, volume~4, pages 25--30 vol.4, June 1992.

\bibitem{extrapolation_and_interpolation}
E.~Barnard and L.F.A. Wessels.
\newblock Extrapolation and interpolation in neural network classifiers.
\newblock {\em IEEE Control Systems Magazine}, 12(5):50--53, October 1992.

\bibitem{PINN}
M.~Raissi, P.~Perdikaris, and G.~E. Karniadakis.
\newblock Physics-informed neural networks: {A} deep learning framework for solving forward and inverse problems involving nonlinear partial differential equations.
\newblock {\em Journal of Computational Physics}, 378:686--707, February 2019.

\bibitem{PINN_extrapolation}
Athanasios Papastathopoulos-Katsaros, Alexandra Stavrianidi, and Zhandong Liu.
\newblock Improving {Physics}-{Informed} {Neural} {Network} {Extrapolation} via {Transfer} {Learning} and {Adaptive} {Activation} {Functions}.
\newblock In Walter Senn, Marcello Sanguineti, Ausra Saudargiene, Igor~V. Tetko, Alessandro E.~P. Villa, Viktor Jirsa, and Yoshua Bengio, editors, {\em Artificial {Neural} {Networks} and {Machine} {Learning} – {ICANN} 2025}, pages 289--301, Cham, 2026. Springer Nature Switzerland.

\bibitem{Badt_Katz_2022}
Noam Badt and Ori Katz.
\newblock Real-time holographic lensless micro-endoscopy through flexible fibers via fiber bundle distal holography.
\newblock {\em Nature Communications}, 13(1):6055, October 2022.

\end{thebibliography}
\end{document}